# A Comprehensive Survey on the Cyber-Security of Smart Grids: Cyber-Attacks, Detection, Countermeasure Techniques, and Future Directions

Tala Talaei Khoei[1], Hadjar Ould Slimane [1], and Naima Kaabouch[1]

[1]School of Electrical Engineering and Computer Science, University of North Dakota, Grand Forks 58202, ND, USA

*Abstract-* One of the significant challenges that smart grid networks face is cyber-security. Several studies have been conducted to highlight those security challenges. However, the majority of these surveys classify attacks based on the security requirements, confidentiality, integrity, and availability, without taking into consideration the accountability requirement. In addition, some of these surveys focused on the Transmission Control Protocol/Internet Protocol (TCP/IP) model, which does not differentiate between the application, session, and presentation and the data link and physical layers of the Open System Interconnection (OSI) model. In this survey paper, we provide a classification of attacks based on the OSI model and discuss in more detail the cyber-attacks that can target the different layers of smart grid networks communication. We also propose new classifications for the detection and countermeasure techniques and describe existing techniques under each category. Finally, we discuss challenges and future research directions.

## I.  Introduction

The traditional power grid is no longer a practical solution for power delivery and distribution due to several shortcomings including: chronic blackouts, energy storage issues, high cost of assets, and high carbon emissions. Briefly, several cases prove that there is a serious need to improve the functionality of the traditional power system. For example, in February 2020, the storm Ciara caused a power cut for around 130,000 homes in France. During the same month, in Bavaria, the storm Sabine caused a blackout for approximately 60,000 homes [1]. In March 2016, at least 70 million people in Turkey were impacted by a power blackout. These examples are the obvious reasons why using a traditional power grid is no longer considered an effective power system [2].

To address the limitations of the traditional grid, a new approach, microgrid, was introduced. A microgrid can be defined as a local and small distribution system that consists of sets of micro sources, namely micro turbines, fuel cells, photovoltaic arrays wind turbines, and some storage systems like energy capacitors. It can be connected to a main grid or work independently. Microgrids provide some benefits, such as a higher efficiency, reduction of emissions, and cheaper and cleaner energy. Also, this technology deals with some challenges, including the resynchronization with the main grid, which can be problematic to the network, due to the network inconsistency. To address these challenges and limitations, a holistic solution, smart grid, was proposed [3]. This new electrical grid includes a variety of operations and energy measures, including smart meters, smart appliances, renewable energy resources, and energy-efficient resources. It utilizes information technology to deliver energy to end-users through a two-way flow of communications, which changes the power infrastructure in terms of efficiency, scalability, reliability, and interoperability.

The National Institute of Standard and Technology (NIST) stated that smart grids consist of seven logical domains, namely bulk generation, transmission, distribution, customer, markets, service provider, and operations. These logical domains have actors and applications that are presented as smart grid's conceptual model. Actors are defined as programs and systems, while applications are considered as tasks. These tasks are conducted by a single or multiple actors in every domain. In the customer domain, the major actor is the end-user, which is divided into three types: home, commercial, and industrial. This domain mainly has a close communication with the distribution, operation, service provider, and market domains. Within the market domain, the users have to be the operators that participate in electricity markets. The service provider consists of the organizations that can provide services to customers and utility companies. More importantly, the bulk generation domain has some electric generators in bulk quantities, and the transmission domain can carry out the generated electric power over long distances from the generation domain to the distribution domain via a variety of substations [3]. The transmission network can monitor and control via Supervisory Control and Data Acquisition (SCADA) system. The distribution domain can distribute the electricity to and from end-users in different structures, like radial, looped, or meshed. This domain is capable of supporting energy generation and storage, which is mainly connected to the transmission domain, customer domain, and metering points for electricity consumption [4].

The smart grid is expected to create a reliable, efficient, and clean energy distribution by combining various technologies. It promises reliability, improved efficiency, and economical means of power transmission and distribution. It also reduces greenhouse emissions to deliver clean, affordable, and efficient energy to users [4]. However, this infrastructure can be subjected to cyber-attacks that can violate the availability, integrity, confidentiality, and accountability of smart grid's security requirements. For example, a cyber-attack on a U.S. power grid occurred in March 2018 that targeted numerous nuclear power plants and water facilities. Another instance of cyber-attacks happened in Ukraine in December 2015. During this incident, the attackers turned off 30 substations that led to a complete blackout for about 6 hours, leaving around 230,000 people without electricity. To improve the security level of the power systems, the US National Electric Sector Cybersecurity Organization (NESCOR) and the Department of Energy (DOE) joined their efforts. For this purpose, they collaborated with some federal U.S. agencies, such as the Cybersecurity for Energy Delivery Systems (CEDS) and the Federal Energy Regulatory Commission (FERC). They involved experts, developers, and users to test the security of the power systems. They collaboratively worked together to enhance security risks of smart and mitigation strategies. Their investigations proved that using this modern technology requires some holistic solutions to defend and prevent cyber-security issues. Despite the critical advancements of smart grids, the detection and prevention of sophisticated cyber-attacks are still at an early stage and need more attention [5, 6].

Over the last decade, several surveys provided an overview of smart grid's cyber-security, as shown in Table I [7, 8, 9-16]. The authors of [7, 8, 9, 11, 14] reviewed the main cyber-attacks that can damage the smart grid infrastructure, the detection techniques, and the countermeasures. In [10, 16], the authors mainly focused on the cyber-physical attacks in smart grid networks, their impacts, along with their defense strategies. In [16], the authors also highlighted a classification for detection techniques for cyber-physical attacks in smart grids and compared the efficiencies of various detection techniques. In [12], the authors reviewed cyber-security related to smart homes and smart grid networks. They classified cyber-attacks in smart home/smart grid networks according to confidentiality, integrity, availability, authorization, and authenticity. The authors of [15] also provided a survey on cyber-security and privacy of smart metering networks. They briefly reviewed cyber-attacks in traditional power systems and smart grid networks and introduced future research trends in depth.

Several other surveys discussed the security of smart grid infrastructure and provided various cyber-attack classifications on smart grid networks. These surveys only classified attacks in terms of the security requirements, such as confidentiality, integrity, and availability. They also did not consider the accountability requirement, which can present any action conducted to guarantee the system traceability. The majority of surveys did not provide a comprehensive overview of detection techniques and countermeasures for smart grid infrastructure. Furthermore, a last 2020 survey related to cyber-security mainly discussed the different cyber-attacks in smart grid in terms of security requirements and four layers of OSI communication model. This survey did not discuss any solutions regarding prevention or defending attacks in the network. Therefore, our study mainly fills the current gap and provides a comprehensive classification for cyber-attacks in the smart grid. This study also provides a new classification for both detection and countermeasure methods, while the previous works did not provide such complete security strategies. To be precise, in this paper, we present an in-depth survey of technological advances in smart grid infrastructure cyber-security. First, we provide a classification of cyber-attacks that target the OSI communication layers. We also propose two classifications, one for detection techniques and another one for countermeasure methods. After each classification, cyber-attacks are described as well as the methods of detecting and mitigating these attacks. We also discuss four critical security parameters for cyber-attacks that can compromise smart grid networks. Moreover, an overview of future trends and the current challenges related to this network's cyber-security are discussed. The contributions of this survey can be summarized in the following key points:

- Review and classification of cyber-attacks in the smart grid network.
- Description and comparison of these attacks, along with their purposes and impacts.
- Review, analysis, and classification of detection techniques.
- Analysis and classification of corresponding countermeasure methods.
- Discussion of the challenges and open issues related to the security of smart grid and future research directions to address them.

The remaining of this paper is organized as follows: Section II provides an overview of the smart grid system and its features and architectures. Section III describes the smart grid architecture, its technologies, and protocols. Section IV reviews the smart grid's security, discusses the security requirements that are expected to be met, and provides a classification of cyber-attacks that target the OSI communication layers. The purposes and impacts of these cyber-attacks are also evaluated in this section. Section IV also provides a classification of detection techniques, the state-of-the art in detection techniques, and summarizes existing countermeasures against various cyber-attacks. Section V describes several research challenges and future research directions. Finally, Section VI closes the survey with a conclusion.

Table I. Existing Surveys Related to the Cyber-security of Smart Grids.

| Related Work | Topic | Cyber-Attacks Mentioned | Concepts Covered | Concepts Not Covered |
|---|---|---|---|---|
| Gunduz et al. [8] | Survey on cyber-security solutions of IOT-based smart grids. | Different kinds of cyber-attacks against CIA tirade and five OSI communication layers. | • Cyber-attack types and the general importance of countermeasures.<br>• Analysis of various cyber-attacks and their security requirements along with future directions. | Countermeasure methods and detection techniques. |
| Peng et al. [9] | Survey on security communications in smart grids. | Traffic analysis, social engineering, scanning I.P., scanning port, scanning vulnerability, worms, DoS, FDI, replay, privacy violation, backdoor | • Cyber-physical security of smart grids, and potential IT-based attacks scenarios.<br>• Detection/protection methods and challenges regarding to threats of smart grids. | Accountability as a security requirement in smart grids. |
| He et al. [10] | Survey on cyber-physical attacks and solutions in smart grids. | Generation system attacks, transmission system attacks, distribution system/ customer side attacks, electricity market attacks | • Critical cyber-physical attacks and their defense methods.<br>• Analyzing the impact of cyber-physical attacks in smart grids. | Detection techniques for cyber-physical attacks in smart grids. |
| Gupta et al.[11] | Survey of cyber-security in smart grids. | DoS/ DDoS attacks | • Smart grid and its components.<br>• Existing methods for communication protocols and their architectures.<br>• DoS/DDoS attacks and their impacts on smart grids. | Existing cyber-attacks that targets smart grids, their countermeasures, and detection techniques. |
| Elmrabet et al. [12] | Comprehensive review of cyber-attacks and their solutions in smart grids. | Traffic analysis, social engineering, scanning IP, scanning port, scanning vulnerability, worms, Trojan horse, DoS, FDI, replay, privacy violation, integrity violation, backdoor, MITM, jamming, popping the HMI, masquerade. | • Important cyber-attacks in smart grid and their impacts.<br>• Various security methods to address cyber-security issues in smart grids. | Detection techniques and countermeasure approaches. |
| Komninos et al. [13] | Survey on cyber-security in smart homes and smart grids. | Different kinds of cyber-attacks against confidentiality, integrity, availability, authorization, authenticity. | • Most common threats against smart homes and smart grids.<br>• Different cyber-attack scenarios with their specific countermeasures.<br>• Methods to defend against or prevent cyber-attacks. | Smart grid cyber-attacks, countermeasures, and detection techniques. |
| Sakhnini et al. [14] | Survey on cyber-security aspects of IOT aided smart grids. | MITM, jamming, FDI, spoofing, DoS, malware, replay attacks. | • Bibliometric analysis of published journals.<br>• Different cyber-attacks targeting smart grids and their security mechanisms.<br>• Future trend of smart grid cyber-security. | Countermeasure techniques. |
| Kumar et al. [15] | Survey of cyber-security and privacy of smart grid metering networks. | Different kinds of attacks on energy companies, renewable energy resources, and metering networks. | • Cyber-attacks vulnerabilities in the traditional energy network.<br>• Security and privacy requirements for smart grid metering networks.<br>• Future research trends and challenges. | Countermeasures and detection methods. |

| Musleh et al. [16] | Survey on detection techniques for cyber-physical attacks in smart grids. | Different kinds of cyber-physical attacks that take place in smart grids. | <ul><li>Cyber-physical attacks and the classification of detection techniques in smart grids.</li><li>Analysis of false data injection attacks and their impacts.</li><li>Study of future trends and their challenges.</li></ul> | Countermeasure techniques. |
|---|---|---|---|---|

## II. Overview of Smart Grid

In this section, we mainly discuss the smart grid's features and applications. In the following, we provide a short description of the most critical features in smart grids and the important applications in this network.

### 2.1. Smart Grids' Features

The significant features expected from the smart grid are improving grid resilience, self-healing, increasing environmental and system performance [7, 17]. Grid resilience means that the power grid can recover quickly and fulfill the mission during power interruptions and outages [18]. This can be provided by adding extra disperse power supply and integrating modern resources into the power grid when an interruption happens [19]. The self-healing feature allows the system to identify faults quickly, decrease the duration of the outage, and help the system to recover faster. Therefore, by providing a higher level of flexibility and reliability, the grid's resilience and self-healing features have a critical impact on the economy [20].

Another expected feature in the smart grid is improving system's performance. In the traditional power grid, energy loss may happen due to several reasons, including faults in power stations or damages in transmission lines. The smart grid promises to increase the system performance by optimizing asset utilization and operations, reducing energy costs, and transmitting electricity in an effective manner. These benefits may directly increase the quality of power and efficient asset management, which indicates the increased level of system performance [20]. Moreover, the smart grid is expected to expedite the replacement of electric vehicles with conventional vehicles. These replacements may lead to enhance environmental performance by reducing the energy used for end-users and decreasing energy loss through the grid [19].

### 2.2. Smart Grid Applications

As illustrated in Fig. 1, the smart grid includes a variety of heterogeneous, distributed applications and capabilities, such as Advanced Metering Infrastructure (AMI), Supervisory Control and Data Acquisition (SCADA), Substation Automation, Electrical Vehicles (E.V.s) charging, Demand Response Management (DRM), Outrage Management (O.M.), Distribution Management (D.M.), and Home Energy Management (HEM) [21, 22]. This section will discuss three vulnerable applications in the smart grid infrastructure, namely AMI, SCADA, and DRM. The other applications were discussed in detail in [21-28].

Advanced Metering Infrastructure (AMI) is one of the essential components of smart grid infrastructure. AMI is mainly responsible for reading the power usage of home appliances and some other integrated devices, such as water heaters, gas meters, smart thermostats, rooftop photovoltaic systems, etc. AMI consists of three main components: a smart meter, a data concentrator, and a central system (AMI headend), with a two-way flow of communications between these components [29]. The meter data that are collected from the power usage of home appliances are received by the AMI host system and transmitted to the meter data management system (MDMS). MDMS is responsible for data storage management and data analysis for the utilities. The AMI system provides financial benefits and increased service quality (multi-utility service and multi-vendor service) [30].

Supervisory control and data acquisition (SCADA) is a type of Process Control System (PCS) that is responsible for monitoring, measuring, and analyzing real-time data of the electrical power grid [7]. SCADA mostly effective for large-scale environments; however, it can ensure both short-range and long-range communications [31]. This system consists of three main elements: The Remote Terminal Unit (RTU), Master Terminal Unit (MTU), and Human-Machine Interface (HMI). RTU, as a device, consists of three components. The first component is the one that can perform data acquisition, the second component runs logic programs that are coming from the MTU, and the third component is mostly responsible for developing the communication infrastructure [32]. Another element in SCADA is the MTU, which is a device for monitoring and controlling the RTU. As the last element in SCADA, HMI considers as a graphic interface for the SCADA operator [22].

Demand Response Management (DRM) is one of the essential systems in the smart grid infrastructure. This system refers to the routines conducted to control the energy consumption of consumers. DRM can achieve a balance between electrical energy supply and demand. DRM's benefits are to decrease the peak-to-average ratio of the demand and power supply, reduce user bills and power generation costs, improve energy efficiency, and address short-term reliability [22].

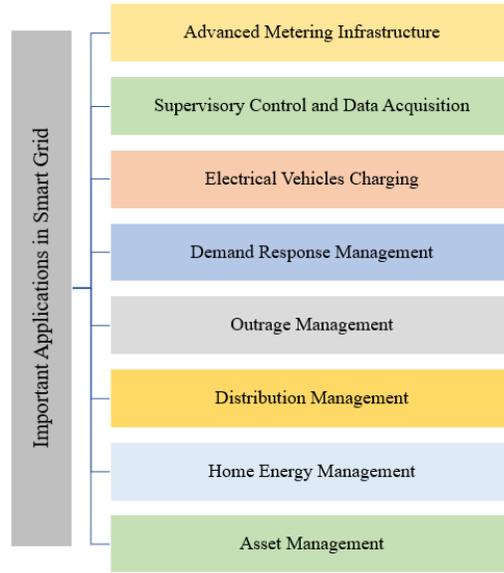

Fig. 1. Important Applications in the Smart Grid.

### III. Smart Grids Architecture

As the smart grid infrastructure connects a huge variety of systems, the hierarchical architecture of the smart grid with few sub-networks is considered critical in the infrastructure; however, each sub-network is only responsible for specific geographical regions. Smart grid network includes three main sub-networks, Wide Area Network (WAN), Neighborhood Area Network (NAN), and Home Area Network (HAN), as shown in Fig. 2 [33]. To these three sub-networks, the authors of [34] added several sub-networks, namely Field Area Network (FAN), Local Area Network (LAN), and Building Area Network (BAN). BAN is divided into two sub-networks, Home Area Network (HAN) and Personal Area Network (PAN), as shown in Fig. 3.

WAN is one of the major networks in the smart grid architecture. In [35], the authors highlighted WAN as the main network that can create a connection backbone to connect highly distributed smaller networks for power systems at various locations. This network is a high-bandwidth connection network, which can deal with long-distance data transmission over advanced monitoring and sensing applications. WAN provides a bidirectional communication for automation, monitoring, and communication of smart grid systems. The authors of [36] described the NAN as a network that is expected to connect smart meters and distribution automation devices to the WAN gateways. It is a bridge between user premises and substations with access points, collectors, and data concentrators. This sub-network may be considered as low bandwidth that is highly robust for secure data communication.

HAN is necessary for customers to monitor and control smart devices and execute some functionalities, such as DRM and AMI. It allows users to know about their electricity consumption cost and handle their usage behaviors. According to [37], this network supports low-bandwidth communication between home appliances and smart meters. In [35], the authors defined BAN as a network that can perform any communications among homes and offices within a building. PAN is responsible for any communication between personal appliances, such as laptops, tablets, phones, etc. In LAN and FAN, any distant communication in backhaul networks, smart homes, factories, or even power generation plants can be performed.

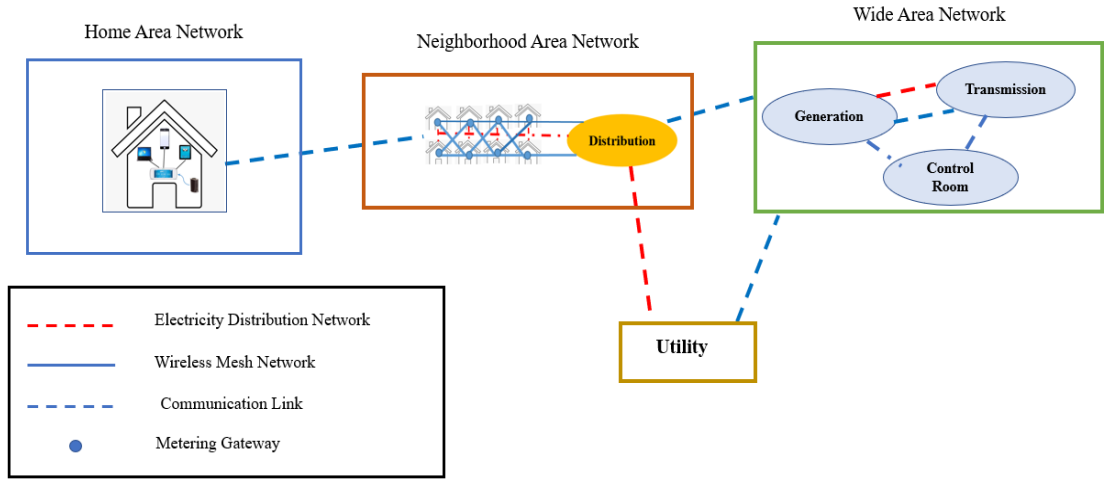

Fig. 2. Main Sub-Networks in the Smart Grid Architecture.

In a sophisticated smart grid architecture, different networks demand various communication technologies and protocols to deliver reliable and secure data or power to utilities and users. In the following sections, we will describe smart grid communication technologies and few numbers of well-known protocols.

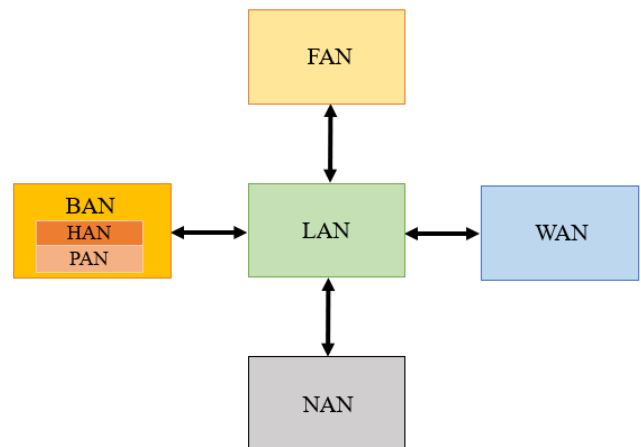

Fig. 3. Different Communication Networks in Smart Grids.

### 3.1. Smart Grids Communication Technologies

In smart grid, secure, reliable, and real-time information is considered a key factor for an efficient delivery of power between generators and users. Equipment failures, natural accidents, catastrophes, and capacity constraints can be the main reasons for power disturbances in grid systems. To deal with these issues, new communication and information technologies with modern intelligent monitoring systems play an indispensable role in securing data transmission between smart meters and utilities, while they apply two different communication media, namely wired and wireless. The authors of [38] discussed the benefits of wired and wireless communication technologies. In this study, the authors highlighted some of the benefits of wireless communications over wired communications, such as reasonable infrastructure prices and stronger connections in unreachable regions; however, this technology is only able to provide a connection in short distances with low data rates, compared to wired communications. Wireless technologies include Zonal Intercommunication Global-standard (Zigbee), Z-wave, worldwide interoperability for microwave access (WiMAX), Wi-Fi, DASH7 (D7A), and cellular and satellite.

Wired communications also have some advances. For example, it can provide higher connection capacity and shorter communication delay with less interference in comparison with wireless communication. Examples of wired communication technologies include Powerline Communication (PLC) and Digital Subscriber Lines (DSL) [39]. In this section, we mainly focus on satellite, PLC, and DSL. For further reading, Zigbee, Z-wave, WiMAX, WiFi, DASH7, and Cellular are discussed in detail in [39-44]. Main features, including bandwidth, coverage rate, data rate, application, and application area of the most common communication technologies along with their standards, are summarized in Table 1. More details about smart grid communication technologies can be found in [38-48].

PLC is a wired communication technology that can support high-speed data from one device to another one. It simply connects smart meters to a data concentrator via a power line, and its data can be transmitted to the data center with cellular network technologies [39]. It is suitable for some applications, including smart metering, home automation, Heating, Ventilation and Air Conditioning (HVAC) control, and lighting. This technology reduces installation costs; however, it has some technical issues, including low bandwidth, high dependency on Quality of Service (QoS), and high sensitivity to disturbances. Because of these limitations, PLC does not provide full connectivity, and it has to be combined with other technical communications, such as General Packet Radio Services (GPRS), and Global System for Mobiles (GSM) [39-41].

Digital Subscriber Lines (DSL) is another wired communication technology that has fast speed digital data transmission and can conduct effectively on the voice telephone network [40]. Although this technology provides cheap, high data bandwidth, and expected availability, it has some disadvantages, such as lack of reliability, downtime, and proper standardization. In [35], the authors also mentioned that as this technology requires cables, it is difficult to use in rural areas due to high installation expenses in low-density regions. Satellite communication is one instance of wireless communication technology that is usually used in radio broadcasting, plane TVs, and vehicles. Although this technology provides some benefits, including reliability and flexibility, its performance is heavily impacted by weather conditions [35], and it is not considered cost-effective. Satellite communication also reduces the need for backhaul networks, which demonstrates a good fit for smart grid infrastructure [43]. This technology can provide SCADA and distant communication remote substations, making it a viable option for future use.

### 3.2. Smart Grids Protocols

Initially, the Transmission Control Protocol/Internet Protocol (TCP/IP) was used in the smart grid to ensured end-to-end data communications. However, this protocol is not considered a good option for smart networks due to its sophisticated memory management problems and the fact that it is only suitable for wide area networks. Alternatively, several different smart grid protocols were developed to meet the different smart network requirements [49].

SCADA and AMI are important key components in smart grid infrastructure. A few protocols were developed over the years to provide secure and reliable communications for these systems. The communication within SCADA depends on several industrial protocols, such as Modicon communication bus (Modbus), Distributed Network Protocol version 3 (DNP3), Process field bus (Profibus), and International Standard Defining Communication Protocol 61850 (IEC61850). However, the communications among the AMI, home appliances, and smart meters are done through various communication protocols. They vary widely in their inherent security requirements and vulnerabilities. In this section, we mainly focus on four vulnerable protocols that are used in smart grid infrastructure, including Modbus, DNP3, Profibus, and IEC61850.

According to [50], DNP3 is an optimized open communication protocol used for power grid equipment. Initially, this protocol's major aim was to be used in the traditional power grid; however, this protocol has recently been used as a solution for delivering data measurements in the smart grid because of its reliability, efficiency, and compatibility in comparison with previous protocol versions. The DNP3 inherently was not a secure protocol; hence the authentication features were added to DNP3 protocol to addresses the security issue. Modicon communication bus or Modbus is another protocol designed in 1979 as a serial communication protocol to permit communication between various machines over twisted wires. This protocol consists of three types, Modbus American Standard Code for Information Interchange (ASCII), Modbus Remote Terminal Unit (RTU), and Modbus Transmission Control Protocol (TCP). In general, Modbus ASCII enables the messages to be coded in hexadecimal, while this is the slowest type of Modbus in comparison with other types, and it is ideal for telephone modems. In Modbus RTU, the messages are expected to be coded in a binary manner. Modbus RTU is suitable to be applied over RS232 and can generate communications between master and slaves by using their IP addresses instead of their device addresses. Modbus TCP is defined as a specific data frame protocol, which has a function code for an action that has to be completed. This protocol is particularly one of the most popular industrial control protocols, which generates a simple request or reply method between the control center and field devices.

Table II. Communication Technologies of Smart Grids.

| Technology | Frequency | Coverage Rate | Data Rate | Application | Application Areas |
|---|---|---|---|---|---|
| ZigBee | 915 MH and 2.4 GHz | 30-100m | Up to 250 Kbps | HAN | Energy Monitoring, Automatic Meter Reading, Home Automation |
| PLC | 24-500 KHz | 1-3Km | 2-3 Mbps | HAN NAN | Automatic Meter Reading, Low Voltage Distribution |
| Bluetooth | 2.402-2.48 GHz | 1-30m | 1 Mbps | HAN | Home Automation |
| Fiber Optic | 100-1000 THz | AON: up to 10 Km BPON: up to 20–60km EPON: up to 20 km | AON:100 Mbps up/down BPON:155–622 Mbps EPON: 1 Gbps | WAN | AMI, Metering reading, Distribution automation, Service switch operation Demand response, Wide-area monitoring |
| WiFi | 2.4 and 5 GHz | Up to 1 Km | Up to 600 Mbps | HAN FAN NAN BAN WAN | AMI |
| WiMAX | 2.3-2.7 and 3.4-3.6 GHz | Almost 10Km-100Km | Up to 75 Mbps | HAN NAN FAN WAN | Wireless Automatic Meter, Reading, Outage Detection, AMI |
| GSM | 850-1900 MHz | 1-10 Km | 14.4 Kbps | HAN | AMI, Demand Response |
| Satellite | 1- 40 GHz | 100–6000 km | Iridium: 2.4–28 kbps Inmarsat-B: 9.6 up to 128 kbps BGAN: up to 1 Mbps | WAN | AMI, Remote generation plants, Electric Vehicles Remote automation, Distribution Automation |
| GPRS | 800-1900 MHz | 1-10 Km | 179 Kbps | HAN | AMI, Demand Response |
| Z-Wave | 868 and 915 MHz | 30-100 m | 40 Kbps | HAN | Home automation, energy automation |
| DSL | 4 KHz-MHz | ADSL: up to 5 Km ADSL2: up to 7 km ADSL2þ: up to 7 km VDSL: up to 1.2 km VDSL2: 300 m–1.5 km | ADSL: 8 Mbps down/1.3 Mbps up ADSL2: 12 Mbps down/ 3.5 Mbps up ADSL2þ: 24 Mbps down/ 3.3 Mbps up VDSL: 52–85 Mbps down/16– 85 Mbps up VDSL2: up to 200 Mbps down/up | HAN NAN WAN | Smart Grid City Smart Metering |
| LTE Mobile Network | 0.41-2.1 GHz | 5-30 Km | 75 Mbps-300 Mbps | HAN | AMI Demand Response |

Process Field Bus (PROFIBUS) is yet another communication protocol in smart grid infrastructure used for automation technology [52]. This protocol is considered as one of the well-known Fieldbus protocols standardized as EN50170. PROFIBUS can address the real-time requirements on MAC layer. It is used as a token-passing protocol, same as IEEE 802.4 in a Token Bus. This protocol is mainly divided into two categories, PROFIBU Decentralized Peripherals (DP) and PROFIBUS Process Automation (PA). PROFIBUS DP is used for conducting sensors and actuators via centralized controllers, and the PROFIBUS PA can be used in hazardous areas, and it is mainly designed as an improvement version of some convenient systems, like Highway Addressable Remote Transducer (HART) in process automation.

Another protocol is IEC 61850 [53], which is another well-known communication protocol in the smart grid mainly designed for communication networks and systems in order to provide better interoperability between Intelligent Electronic Devices (IEDs). It provides several opportunities to increase the efficiency of the grid and reduce its cost. This protocol can introduce five variant types of communication services, including Abstract Communication Service Interface (ACSI), Generic Object-Oriented Substation Event (GOOSE), Generic Substation Status Event (GSSE), Sampled Measured Value multicast (SMV), and Time Synchronization (T.S.).

## IV. Security of Smart Grids

With the transformation of traditional power grids to smart grids, the security became one of the critical challenges in the last few decades. To address this challenge, the system and its infrastructure must be designed following secure architectural conditions. Therefore, cyber-security as an integral and complimentary process needs to follow a set of comprehensive security requirements. Initially, the National Institute of Standards and Technology (NIST) defined three security requirements that need to be met in the smart grid: confidentiality, integrity, and availability. However, the authors of [54] demonstrated that accountability also plays an important role in the security of smart grids.

In general, when unauthorized access to some information happens, confidentiality is lost. While integrity seeks to deliver accurate data by protecting it from any improper modification or data destruction done by an unauthorized user. Availability, on the other hand, is defined as an important aspect of smart grids that can guarantee access to the system's data. Loss of availability indicates that the data is not available or accessible to use by users. In addition to the requirements previously mentioned, accountability plays an important role in smart grid security; it guarantees the system's traceability that must be recorded by a person, device, or public authority. Moreover, the recorded data can be used as an evidence in case of an attack to prove the action made by every user, or even administrator, and the integrity of the data collected from each device [5]. Hence, following these four requirements, including confidentiality, integrity, availability, and accountability can provide adequate protections to smart grid infrastructures.

Due to the inherent vulnerabilities of communication, smart grid networks are subject to several cyber-attacks, which can be classified in different ways. In the following section, we discuss existing cyber-attack classifications along with our proposed classification, and describe the possible attacks, with their purposes and impacts on the networks.

### 4.1. A Classification of Cyber-Attacks in Smart Grid

Fig. 4 provides a visual representation of existing cyber-attack classifications in smart grids in the current Fig. 4 provides a schematic of existing cyber-attack classifications in smart grid. In [55, 56], the authors classified these cyber-attacks based on the security requirements, confidentiality, integrity, and availability; however, they excluded accountability from this classification [7, 54]. As shown in Fig. 4, the authors of [56] also provided another cyber-attack classification, which was according to the subnetworks and architecture of smart grids, namely home area network, wide area network, and neighborhood area network. This classification does not include cyber-attacks on the other sub-networks, such as Building Area Network (BAN), Field Area Network (FAN), and Personal Area Networks (PAN). In this paper, the authors also described the impact of each cyber-attack on three security requirements (confidentiality, integrity, and availability), while they excluded the attacks targeting accountability.

In [7], the authors proposed a classification based upon the attacking cycle, including reconnaissance, scanning, exploitation, and maintenance access, as shown in Fig. 4. This classification also did not include all cyber-attacks. Several cyber-attacks, such as intrusion, brute-force, and spoofing attacks, are primary concerns in a smart grid, and they require multiple security mechanisms. However, this paper excluded these cyber-attacks from the classification. As illustrated in Fig. 4, the authors in [58] also provided a general classification divided into three categories: component-wise, protocol-wise, and topology-wise. The authors indicated that some cyber-attacks such as replay attacks and eavesdropping attacks might be excluded from this classification. The authors of [59] proposed a taxonomy of basic cyber-attacks, including devices, data, privacy, and network availability attacks. This classification also does not include cyber-attacks like social engineering attacks. In [8], the authors classified cyber-attacks according to five communication layers, including the application layer, transport layer, network layer, MAC layer, and physical layer. Several attacks target the session layer and presentation layer, which both were excluded from the classification. This paper also did not provide clear descriptions of cyber-attacks, with their impacts, purposes, and security requirements they target, and did not cover the detection and mitigation techniques.

A number of survey papers related to cyber-attacks on smart grid networks have been published over the last decade. Some of these papers focus on cyber-attacks that target one or some of the communication layers, such as the physical layer or the network layer. For example, a survey published in 2020 focused on the attacks based on the layers of the TCP/IP model; however, the TCP/IP model does not distinguish between the cyber-attacks that target the application, presentation or session layer and the data link or physical layer. Therefore, motivated by the limitations of the current studies, we classify cyber-attacks on smart grids based on the seven communication layers of the OSI model, which provides a comprehensive conceptual detail of the networking process. The seven layers of this model are introduced to perform a set of unique functions in a data communication. As a result, the OSI model is more detailed and informative compared to the TCP/IP model. Since several cyber-attacks may target these layers, it is important to select a model that considers a set of distinct functions for every layer. Therefore, we classify the cyber-attacks in smart grids into the physical, data-link, network, transport, session, presentation, and application layers.

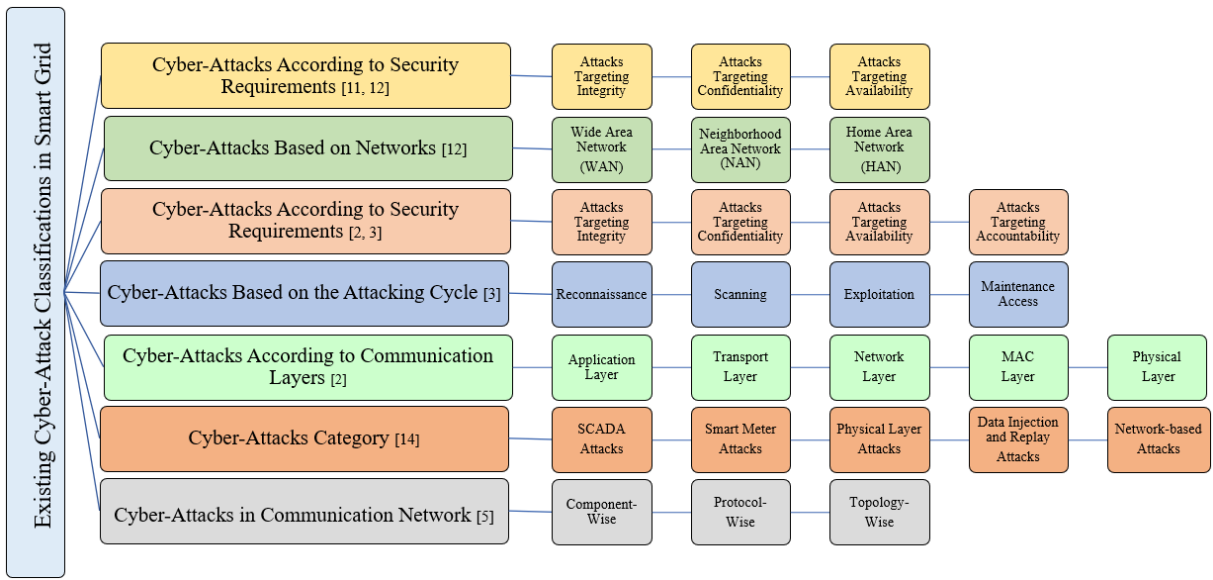

Fig. 4. A Review of the Current Classifications of Cyber-Attacks in Smart Grids.

### 4.2. Cyber-Attacks in Smart Grids

As we mentioned, multiple cyber-attacks may target smart grid infrastructures. Fig. 5 illustrates possible cyber-attacks in a smart grid, with their target layers in OSI model. According to this figure, multiple cyber-attacks can target the same communication layers or target more than one layer simultaneously. In the following, we describe these cyber-attacks, along with their purposes, targeted layers' impacts, and their security requirements, as summarized in Table III.

One of the cyber-attacks that are more likely to occur in a smart grid is jamming. In these attacks, an attacker broadcasts continuous or random signals to keep the channel busy and prevent authorized devices from transmitting and receiving [7, 59, 67]. Different types of jammers include constant, random, deceptive, and reactive jammers

[64] can target the physical layer [55, 8, 61], data-link layer [8], and network layer [87] of the smart grid, which can compromise the availability of the smart grid network [7, 8,57].

Spoofing attacks are yet another category of cyber-attacks that target smart grid networks. This category includes identity/data spoofing, Address Resolution Protocol (ARP) spoofing, Global Position System (GPS) spoofing, IP spoofing, and Media Access Control (MAC) spoofing. In any of these attacks, the spoofer pretends to be a legitimate node [7, 8, 57] to mislead other nodes in the network in order to disrupt the security, reliability, stability, and operation of the network [8], which can violate the integrity, the confidentiality and the accountability of the smart grid [8, 57]. These attacks can target the physical layer [71, 72], the data link layer [2, 8, 73], and the network layer [70].

According to the authors of [89], injection attacks can be conducted when an adversary attempts to delete, alter and add new manipulated data to the network, which may disrupt the smart grid operations and lead to a blackout. Violation of the data integrity, corruption, and appearance of illegitimate nodes in the network are also considered as other impacts of this type of cyber-attacks. Like the above-described attacks, injection attacks can target one or several communication layers, such as the data-link layer [8, 72], network layer [8, 72], and transport layer [8, 72]. Another potential cyber-attack that targets smart grid networks, as illustrated in Fig. 5, is the flooding attack which aims to flood the network with several random packets and requests. This attack can occur in the application layer [8, 74] and the network layer [57] to disrupt the system's availability. It may exhaust the target's resources by processing the received fake messages [74]. Another impact of this attack is the lack of node functionality in the network [73].

Other cyber-attacks on smart grid infrastructures are the Man-in-the-Middle (MITM) attacks. Such cyber-attacks target the network layer [8, 13] and the session layer [76]. MITM in a smart grid is performed when an adversary intercepts the traffic by inserting himself between two authorized devices, connecting to the devices, and relaying the traffic between them [7, 10]. The devices seem to communicate directly; however, the adversary is communicating with these devices as a third device. [7, 8] The main purpose of this kind of cyber-attacks is to prevent network data from being transmitted, modify it during the transmission, and gain unauthorized access to valuable data [7, 26]. MITM also can compromise the confidentiality and integrity of the network [57, 75].

Other possible cyber-attacks on smart grid infrastructure are social engineering attacks. These attacks target the application layer and violate the confidentiality of the system [77, 78]. In [78], the authors state that social engineering attacks are considered the biggest cyber-security threat. They described several types of social engineering attacks, including phishing, pretexting, baiting, tailgating, ransomware, fake software, reverse social engineering, phone/windows fraud, and robocalls attacks. All these attacks aim at manipulating users in order to discover and steal valuable and sensitive information. Violating consumers' privacy, identity theft, and stealing sensitive information are the consequences of these attacks.

Eavesdropping attack is another well-known passive attack on smart grid communication channels that targets the network layer [8, 81] and compromises the confidentiality security requirement of the smart grid [8]. In [57], the authors explained that eavesdropping attacks occur when a malicious user listens to the communication between two nodes on a LAN network and gains access to some information. The malicious user may also use this private data to disrupt or compromise the network [8]. These attacks violate the privacy requirement of networks [57, 81].

Time Synchronization Attacks (TSA) are well-known potential cyber-attacks on a smart grid that target timing information [7, 86] at the physical layer [8] and data link layer [8]. TSA can target phasor measurement units and wide area protection, monitoring, and control [57, 8]. The authors of [88] provided a detailed overview of TSA impacts on smart grids. In a smart grid, several applications use synchronous measurements, and the majority of the measurement devices are equipped with GPS for accurate timing information. These devices can also be subject to GPS spoofing attacks. Since the communication and control messages are time-sensitive, GPS spoofing and TSA can be both amongst cyber-attacks that can more likely be carried out in smart grids [7].

Brute-force attacks consist of hybrid brute-force (dictionary attacks), reverse brute-force, and credential stuffing attacks that target the presentation layer, session layer, and network layer [74]. These attacks can occur when malicious attackers crack passwords or passphrases to access the user's accounts or systems. The authors of [67] highlighted the consequences of these attacks, including gaining unauthorized access to the system and user accounts and exploiting the security of the system by compromising the confidentiality and the integrity of the system. In smart grids, an attacker usually benefits from brute-force attacks by gaining access to the private information of consumers in the network [65].

Another cyber-attack against smart grid is the intrusion attack, in which an adversary exploits the vulnerabilities of the network to gain illegal access to the nodes. In other words, any unauthorized or even forcible action may subject to an intrusion attack [75]. It also aims at misusing the available resources in the network by disrupting the integrity, and the confidentiality of the network [82] in both the application layer [83] and the network layer [82]. Due to the smart grid's vulnerable critical nature, the intrusion attack plays an essential role in security disruptions in the network. For example, modern SCADA systems in smart grids experience a lack of authentication and integrity, which causes them to be more exposed to cyber-attacks, such as intrusion attacks. Therefore, the detection and prevention of this attack can improve the network's general performance and avoid system disruptions.

Traffic analysis attack is applied when an adversary listens and analyzes the traffic. The goal of this attack is to control the hosts and devices that are connected to the smart grid network. This attack can violate the confidentiality of the network and target the data link layer. In this attack, the intruder can sniff and analyze the messages, therefore getting information about the patterns of communication between nodes. Masquerading attack is also another known cyber-attack that targets the data link layer in the smart grid. This attack mainly compromises the confidentiality, integrity, availability, and accountability of the network. In such attack, a malicious user may pretend to be an authorized user in order to gain access to the network or be able to conduct unauthorized actions. In the smart grid, the attacker mostly changes a Programmable Communicating Thermostat (PCT) in order to decrease electronic power at a residential location [2, 3].

In the smart grid, one of the most common attacks is the smart meter tampering attack. It can violate the integrity of the network while it targets the physical layer. In smart meter tampering attack, the intruder can modify the transmitted data for any customers. As a result, the user may need to pay higher or lower electricity bills. One cyber-attack that is more likely to happen in the smart grid is known as buffer overflow, in which the malicious attacker sends data to specific components or systems. It also targets the application and transport layers, while it compromises the availability requirement of the network. This attack may lead to system crash and exhausting the network resources [2].

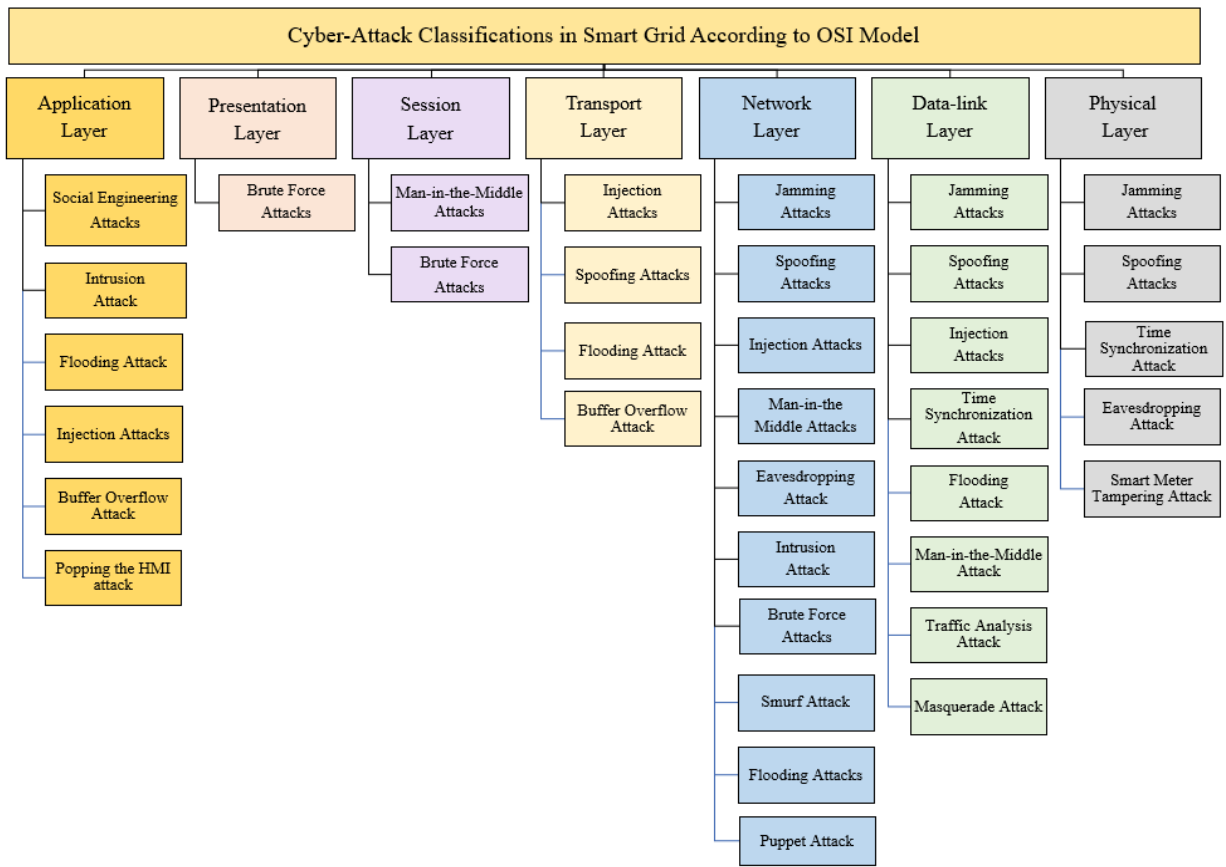

Fig. 5. Cyber-Attack Classification Based on Communication Layers.

Table III. Cyber-Attacks in Smart Grids.

| Cyber-Attacks | Objectives/Purpose | Targeting Layers | Impacts | Security Requirements |
|---|---|---|---|---|
| Jamming Attacks | Disrupting the transmission and the reception of data. | Physical Data Link Network | Blocking one or several nodes to transmit and receive information collisions. | Availability |
| Spoofing Attacks | Pretending to be a legitimate node to compromise the system. | Physical Data Link Network Transport | Misleading other nodes. | Integrity Availability Confidentiality Accountability |
| Injection Attacks | Injecting false/untrusted data packets into a network. | Data Link Network Transport Application | Injecting false data Corrupting the legitimate processes and operations Appearance of illegitimate nodes in the network. | Integrity |
| Flooding Attack | Depleting, and exhausting system resources. | Data Link Network Transport Application | Malfunction of nodes and loss of availability in a network. | Availability |
| Man-in-the-Middle Attacks | Preventing, or modifying data during transmission through the network. | Data Link Network Session | Unauthorized access to sensitive information. | Integrity Confidentiality |
| Social Engineering Attacks | Manipulating users to reveal sensitive information. | Application | Violation of users' privacy. Temporary or permanent damage to the system. Steal sensitive and private information. Identity theft. | Confidentiality |
| Eavesdropping Attack | Monitoring and capturing all network traffic. | Physical Network | Loss of privacy. | Confidentiality |
| Intrusion Attack | Gain illegal access to the node or network. | Network Application | Misusing available resources in the network. | Integrity Confidentiality |
| Brute Force Attacks | Cracking usernames and passwords. | Network Session Presentation | Gaining unauthorized access to users' system or accounts. | Integrity Confidentiality |
| Time synchronization Attack | Targeting timing data and disrupting the time synchronization between nodes. | Physical Data Link | Compromising events, such as location estimation and fault detection Performance degradation. | Integrity Availability |
| Traffic Analysis Attack | Control the hosts and the devices that are connected to the network. | Data Link | Sniff and analyze the message in order to achieve information about the patterns of communications between nodes. | Confidentiality |
| Masquerade Attack | Pretend to be an authorized user. | Data Link | Gaining unauthorized access to users' system. | Integrity Availability Confidentiality Accountability |
| Smart Meter Tampering Attack | Modification the transmitted data for any customers. | Physical | Pay higher or lower electricity bills. | Integrity |
| Buffer Overflow Attack | Sending improper or incorrect data to the specific system. | Transport Application | System crash or exhaust resources. | Availability |
| Puppet Attack | Sending fake data in the AMI network. | Network | Reduce packet delivery to 10% or 20% Exhaust the communication network bandwidth. | Availability |
| Teardrop Attack | Modification of the length and the fragmentation offset in sequential IP packets. | Network | System crash. | Availability |
| Smurf Attack | Modifying the traffic of an entire system. | Network | Replay and saturate the target network. | Availability |
| Popping the HMI Attack | Get unauthorized access | Application | Controling the compromised system | Integrity Availability Confidentiality Accountability |

Another known attack that targets the smart grid is the puppet attack, which violates the availability of the network and targets the network layer. This attack targets the AMI network in the smart grid, using a vulnerability in the Dynamic Source Routing (DSR) protocol. Then, it can exhaust the communication network bandwidth. One of the main impacts of this attack is the reduction of the packet delivery by 10% or 20% [3]. In addition, the smurf attack is one of the potential cyber-attacks in the smart grid that violates the availability of the network. This attack can not only target a specific unit of the smart grid, but also saturates and congests the traffic of an entire system. This attack consists of three factors, namely the source site, bounce site, and target site. In the source site, an adversary sends some spoofed packets to the broadcast address of the bounce site. As soon as the bounce site receives the forged packets, it can broadcast these packets to all hosts. This process may lead to saturate the target network. This attack type mostly targets the network layer [3].

Popping the HMI attack is one of the disruptive cyber-attacks targeting the smart grid. In this case, an adversary uses a common devices's attack (device's software or operating system vulnerabilities) and installs a remote shell, which permits the attacker to connect remotely to the server from the attacker's computer. The aim of this attack is to get unauthorized access and be able to control the compromised system. SCADA and substations of smart grid are considered good targets for this attack. Because the devices' documentaries are publicly available, this attack does not need any advanced networking skills. Therefore, launching such attack is easy and provides full control of the target system to the attacker. It violates the availability, integrity, confidentiality, and accountability and targets the application layer [3].

### 4.3. Detection Techniques of Cyber-Attacks on Smart Grids

Techniques to detect cyber-attacks that target smart grids can be mainly classified into six categories: localization-based techniques, AI-based techniques, prediction models, Channel characteristic-based techniques, filtering-based techniques, and intrusion detection systems.

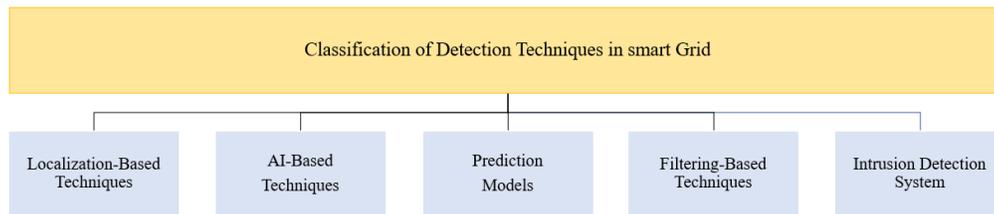

Fig. 6. Classification of Detection Techniques in Smart Grids.

#### 4.3.1. Localization-Based Techniques

Several localizations or ranging techniques have been proposed in the literature and received considerable attention from researchers. In [90], the authors divided the localization techniques into Received Signal Strength (RSS-based), Received Signal Strength Indicator (RSSI-based), Time Difference of Arrival (TDoA-based), and Angle of Arrival-based localization (AoA-based).

RSS signals are widely used in communication technologies for various purposes. One information that can be derived from RSS is the transmitter's location, which has been the focus of numerous studies. For instance, the authors of [91] proposed an RSS-based technique for detecting spoofing attacks based on a spatial correlation feature. In this study, the authors extracted RSS stream features in order to decrease the redundancy of data and applied two distinguishable features of RSS streams, including the Summation of Detailed Coefficient (SDCs) in Discrete Haar Wavelet Transform (DHWT) and the ration of out-of-bound frames. Their proposed approach provides an effective, low-cost method for detecting spoofing attacks in a network. However, the authors of [92] highlighted that the RSS-based technique suffers from poor localization accuracy, which is a critical disadvantage of this technique. It is one of the simplest localization techniques that consider a cost-effective method in networks.

In [93], the authors proposed RSSI-based detection mechanism for MITM attacks. In this study, RSSI values are commonly arbitrary integers, which are received by antennas. These values can be evaluated using a sliding window, that leads to static information about signals' characteristics, including mean and standard deviation. Their proposed technique can detect MITM attacks by analyzing these profiles. In [94], the authors also took advantage of RSSI-based techniques for detecting spoofing attacks in smart grids. They developed a cosine-similarity method of RSSI in home area networks in a smart grid. Their proposed technique provided a higher detection rate compared to other related studies and proved an efficient approach in terms of accuracy [95- 101]. The authors of [102] combined RSS-based technique with the maximum likelihood estimation to handle uncertainty in measurements. According to their simulation results, this technique outperforms the existing RSS-based techniques in detecting attackers.

Localization-based techniques also consist of another common type, TDoA-based. Methods under this type are well-known techniques that measure the distance between nodes. For instance, the authors of [103] introduced a lightweight TDoA-based technique between source and beacon nodes. The authors of [104] also proposed a TDoA-based technique to detect time synchronization attacks in a network. In this work, the authors focused on using TDoA-based techniques on fixed sensors whose time reference could be maliciously affected. Their proposed solution mainly exploits the weighted least squares estimator with newly generated weights and the measurements of TDoA conducted from an unknown source. In [105], the authors developed a TDoA-based technique combined with the Maximum Likelihood Estimator (MLE), which provided superior performance compared to some other existing approaches.

In [106], the authors introduced a new localization technique, AoA-based technique. This technique mainly focuses on the angle of arrival signals to calculate transmitters' locations. In [107], the authors discussed this localization technique, which can achieve the angle of data by using radio array approaches. They highlighted two ways to evaluate the angles of arrival, multiple and directional antennas. Multiple antennas work based on time analysis or even phase difference between the signals at various array elements in which the locations are known. While, directional antennas can compute the RSS ratio between several directional antennas in order to have an overlap between their major beams. In [108], the authors demonstrated that AoA-based localization techniques are not a good fit for cyber-attack detection of an indoor system in networks. In this study, the system's accuracy was reduced because of intensive multi-path components and Non-Line of Sight (NLOS) communications.

### 4.3.2. Artificial Intelligence-Based Techniques

AI-based techniques category consists of various machine-learning and deep learning algorithms, data mining techniques, evolutionary algorithms, and fuzzy logic methods. For detecting cyber-attacks, machine learning category received more attention from researchers. For example, the authors of [55] used machine-learning to detect jamming attacks, namely random forest, support vector machine, and neural network. Their numerical results show that the proposed technique based on random forest achieves high accuracy. In [109], the authors also used machine learning algorithms to detect social engineering attacks. The technique performs based on unsupervised learning, which means that there is no previous knowledge about the observed cyber-attacks. They compared the performance of different machine learning algorithms (support vector machine, biased support vector machine, artificial neutral, scaled conjugate gradient, and self-organizing map) in terms of reliability, accuracy, and speed. Their simulations proved that support vector machine give better results compared to other algorithms.

In [81], the authors used machine-learning techniques to detect brute force attacks on the Secure Shell protocol (SSH) at the network layer. The authors used different classifiers, including K-Nearest Neighbors (KNN), decision tree, and Naïve Bayes (N.B.), to develop scalable detection models that can provide good prediction results. Another study in the literature that exploited machine-learning is detailed in [110]. In this study, the authors highlighted a concept from statics and economics, named "first difference," which led them to develop a classifier to detect time synchronization attacks in the network. Their results show that Artificial Neural Networks (ANN) are the best choice for detecting these attacks in the network. In [111], the authors used an ANN model to detect MITM attacks and their results did provide a good detection rate. In [112], the authors used machine learning techniques to detect and locate intruders in smart grids. The simulation results of this study showed that the proposed method could achieve a good detection rate.

Deep learning techniques has also been used to detect cyber-attacks targeting smart grid infrastructure. For instance, in [113], the authors proposed ensemble deep learning techniques, using deep neural network (DNN) and decision tree. The proposed model was evaluated based on the 10-fold cross validation. The evaluation results showed that the proposed model outperforms other traditional techniques, including random forest, AdaBoost, and DNN. In [114], the authors applied a deep reinforcement learning based technique to identify the physical tripped line and the fake outrage line. Another study [115] also employed a deep learning technique, called encoders to reduce dimensions and feature extraction, followed by an advanced Generative Adversarial Network (GAN) to detect false data injection attacks. In this study, due to expensive costs of labelling in power systems, the collected data is partially labelled, therefore, numerical results show the effectiveness and high rate of accuracy of this model.

Another type of AI-based category is data mining algorithms that can be useful for detecting cyber-attacks in the smart grid. The authors of [16] surveyed existing studies that used data mining techniques for detecting false data injection attacks (FDIA) in the smart grid. These techniques can determine patterns in huge datasets in order to analyze invisible patterns of data. The authors of [111, 112] used a data mining method, Common Path Mining (CPM), to detect FDIA in networks. They introduced a path as a sequence of samples in a temporal order. For any unique event, there is a path, which consists of various types of faults. Hence, when a sequence is compatible with the paths, it will be listed as an attack. In [112], the authors also introduced a Casual Event Graph (CEG) to detect FDIA in smart grids. The main objective of data mining techniques in this study is to train historical datasets. Although data-mining techniques provide some benefits, they may sometimes require low computational complexity (based on the data size) when a training process is over, which is considered a benefit in detecting FDIA in a smart grid.

Fuzzy logic-based methods have also been proposed to effectively detect various attacks in a network. For example, the authors of [116] proposed artificial immune systems and fuzzy logic in order to detect flooding attacks in a network. In this study, the aim of using fuzzy logic is to reduce uncertainty whenever there is no clear line between malicious and legitimate traffic. Another study that applied fuzzy logic in cyber-attack detection is described in [117], in which the authors developed a detection technique based on fuzzy logic for jamming attacks. This technique uses the clear channel assessment, bad packet ratio, and received strength signal parameters to detect link loss due to jamming or other causes. The efficiency of their proposed techniques for constant and random jamming was high. Other authors [118] combined fuzzy logic with other approaches, such as genetic algorithms and Hidden Markov Model (HMM), to detect various cyber-attacks.

Another important type in AI-based techniques is that of evolutionary algorithms, which are widely used for global optimizations. One popular instance of evolutionary algorithms is genetic algorithms. These algorithms can mimic the evolution and natural selection process. In [119], the authors proposed a technique based on a genetic algorithm that consists of two stages, training and detection. In their work, they applied a genetic algorithm for reducing the set of features in the detection stage. According to the authors' results, this technique provides a high level of accuracy for various cyber-attacks in networks. In [120], the authors also analyzed the impact of genetic algorithms on various machine-learning techniques, such as SVM, KNN, and ANN. The simulation results show that genetic algorithms and these three machine learning techniques effectively detect FDIA in smart grids. However, KNN and SVM were found more efficient in detecting these attacks than some existing techniques.

In another work, the authors proposed a hybrid technique based on Genetic algorithms and fuzzy logic [121]. They developed a multi-objective genetic-fuzzy model for detecting anomalies in networks. The numerical results show that the proposed method is not suitable for detecting attacks in networks. In [122], the authors also analyzed a hybrid framework for detecting different cyber-attacks that apply both genetic and fuzzy logic techniques. This method provided good accuracy and better results compared to some other existing techniques.

### 4.3.3. Prediction Models

With the increasing variety and number of cyber-attacks in smart grids, it has become challenging to detect cyber-attacks in any network. The process of detecting attacks usually occurs late for a victims' network. Therefore, detection and identification of attacks in an early stage are considered a challenge for modern systems. Prediction models are well-adapted methods to predict attacks at an early stage in the systems. They mainly apply statistics for the prediction of the results of any unknown events. Several studies focused on using prediction models for attack detections in smart grid infrastructures. For example, the authors of [123] proposed a detection method that uses cosine similarity and chi-square detector to identify FDIA in networks. They also employed Kalman filter to find expected measurements and calculate any deviation between actual measurements and estimated values. Their results show that both chi-square detector and cosine similarity machining are effective methods for the detection of random attacks. In addition, the authors concluded that chi-square detector cannot detect FDIA based on their methodology; however, using cosine similarity provided better results in detecting FDIA in networks.

In another paper, Kalman filter was used to improve cyber-attack detection performance [124]. They modeled the smart grid network as a discrete linear dynamic system and exploited Kalman filter as the state estimation. Several studies also used Kalman filter as a technique to enhance cyber-attack detection in a smart grid, along with other techniques, such as the Euclidean detector [125], cosine similarity detector [126], and chi-square detector [120] which have the capacity of testing the actual measurements and the predicted measurements. Auto Regressive Moving Average (ARMA) and Auto Regressive Integrated Moving Average (ARIMA) are other instances of prediction models that have been proposed for detecting cyber-attacks in networks. In fact, auto regression-based models can predict future trends from past behavior, while moving averages can predict long-term behaviors. ARIMA is also another statistical model that uses time-series data to forecast future behaviors.

In [128], the authors developed an early-stage method to detect SYN flooding attacks. In this method, the SYN traffic is predicted by using ARIMA model, and a cumulative sum algorithm is used to discover SYN flooding attacks. In [158], the author discussed using principal component analysis, ARMA, and Hopefield model to detect intrusions in networks using data mining techniques. In this work, analysis time series data analysis was conducted using these models and neural networks for detecting intrusions in real-time. Their proposed model provided a good detection rate and a low false alarm ratio. Other studies also used these time series models for detecting cyber-attacks smart grid networks [129-134].

### 4.3.4. Filtering-Based Techniques

Filtering-based techniques represent another common category for cyber-attack detections in a smart grid. This survey will discuss two main techniques in this category, namely threshold-based and bloom filtering techniques. Several studies evaluated the efficiency of threshold-based techniques in detecting cyber-attacks in any system. For example, the authors of [103] used threshold-based techniques to detect social engineering attacks in networks. They are easy to develop but not efficient due to their values' limitations. On the other hand, the authors of [135] used bloom filters to detect flooding attacks against signaling protocols. They also introduced a metric called session distance to detect flooding attacks. In addition, they also used the bloom filters in the SCADA system. Because these filters need low memory and computing power, they can effectively help detect any existing anomalies in the SCADA system.

In [136], the authors used a hybrid model for detecting intrusion attacks in SCADA systems. They proposed an approach using multi-level methods to detect anomalies using bloom filters in SCADA networks. They also suggested an algorithm for secure feature extraction and multi-level anomaly detection. Their experimental results showed that the proposed approach can achieve an accuracy of 97%. In [81], the authors compared filtering techniques with some other detecting techniques to detect some social engineering attacks. In this work, the authors highlighted filtering techniques as easy techniques to use, but ineffective and costly. Despite their limitations, these techniques, particularly bloom filters, are known as space efficiency techniques, which are useful in specific scenarios in a smart grid.

### 4.3.5. Intrusion Detection Systems

Intrusion detection systems (IDS) are considered as one of the main techniques to detect cyber-attacks in smart grid infrastructure [183, 184]. These systems can audit and analyze security information to detect any possible malicious vulnerabilities. One of the important benefits of these systems is to detect unknown or zero-day attacks effectively [137, 185].

Several studies have been proposed to detect cyber-attacks using IDS. For example, the authors of [138] proposed a hierarchical distributed IDS based on a distributed fog architecture. This system consists of three different layers of architecture, namely home area networks, residential area networks, and fog operation center networks. Their proposed system demonstrated good performance results over different conditions of the smart grid infrastructure.

In [137], the authors proposed an IDS system to detect operational data. For this purpose, they used real power plant data and described a new architecture for the proposed system. Their simulation results proved that this system has some benefits compared to other existing systems. In [139], the authors proposed a network-based IDS system based on a moving target defense technique in the smart grid. In this study, the authors mainly focused on IPV6 advanced metering infrastructure. The authors of [140] also developed an IDS system, called ARIES, which is able to detect any cyber-attacks, such as DoS, brute-force, port scanning, and bots attacks, against network flows, Modbus/Transmission Control Protocol (TCP), and operational data. They highlighted that the proposed system provides a high efficiency in detecting cyber-attacks. In addition,, some works primarily attempted to improve signature-based IDS systems. In [141], the authors compared different types of IDS systems, including anomaly-based and signature-based. In particular, the authors focused on the improvement of signature-based IDS. To address this challenge, they employed Snort using a layered dataset..

In [142], the authors designed a signature-based IDS system that can detect DoS attacks in a network. For this purpose, they simulated different types of DoS attacks, such as Hello flooding attacks using Cooja simulator and

IPV6 routing (RPL) protocol. Their proposed system provided effective results. In addition to these traditional methods, some studies used hybrid methods that combine IDS with other techniques. For instance, in [143], the authors applied a hybrid model to detect cyber-attacks. This model combines AI-based algorithms, including decision tree, K Nearest Neighbor (KNN), and Support Vector Machine (SVM), along with an IDS system to improve the performance of this system. Their results showed that their proposed system achieves high performance results. The authors of [144] proposed an IDS system that is capable of detecting lethal. The proposed system uses the Cumulative Sum (CUSUM) algorithm with the characteristics of IDS systems. Their achieved detection rate is high, while the false positive rate is relatively low.

### 4.4. Countermeasures in Smart Grids

Several countermeasures have been proposed in the literature that can be used against various cyber-attacks introduced in the next section. For example, in [63], the authors surveyed several countermeasures, including frequency hopping spread spectrum for jamming attacks in wireless sensor networks (WSNs). In [145], the authors compared several encryption techniques and showed that one-time pad (OTP) is the only secure cryptosystem countermeasure solution for brute-force attacks. According to the shift-invariant feature of the transmission policy, the authors of [146] proposed a countermeasure technique for time synchronization attacks. This technique can construct a shift-invariant transmission policies by characterizing the lower and upper bounds for the system estimation, while the attacker does not have any knowledge of the system.

Several studies provided countermeasure classifications. For instance, in [97], the authors proposed a countermeasure classification comprised of four categories, cryptographic functions, personal identification, classification algorithms, and channel characteristics. In [55], the authors divided cyber-attack countermeasures into two categories, cryptographic and network countermeasures. In this section, as Fig. 7 illustrates, we classify countermeasure techniques in smart grid networks into two main categories, computer-based and non-computer-based. Computer-based countermeasures are classified into five types, namely secure protocols and standards, cryptographic functions, intrusion preventions, spread spectrum techniques, and game theory-based techniques. Non-computer-based countermeasures include two types, education, and access control and cyber-security policies. Relevant countermeasures techniques and cyber-security strategies for each category are described below.

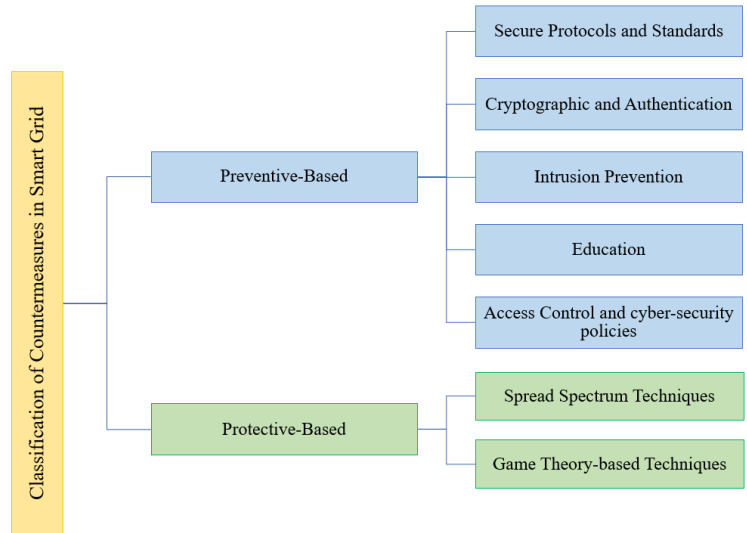

Fig. 7. A Classification of Countermeasure Techniques in Smart Grids.

#### 4.4.1. Preventive-Based Countermeasures

To be prepared for potential cyber-attacks in smart grid network, it is important to understand the different categories of countermeasures. This section only considers computer-based countermeasures that mainly focus on Information technology and software aspects of countermeasures. We will briefly discuss each of these categories with several suggested solutions for smart grid.

##### 4.4.1.1. Secure Protocols and Standards

Secure protocols, such as IPsec, transport layer security (TLS), secure sockets layer (SSL), and secure DNP3, play critical roles in data transmission's security and integrity in smart grid network. However, DNP3 and secure

DNP3 are the most widely used protocols as industrial protocols without any other security mechanism [7]. The authors of [55] suggested these protocols for dealing with several attacks, such as man-in-the-middle, jamming, and eavesdropping attacks in networks.

In [148], the IEEE 802.1 1i protocol was proposed for more confidentiality, integrity, authentication, and availability in a smart grid network. This protocol was designed to replace the Wired Equivalent Privacy (WEP) in the original IEEE 802.11 with AES- CCM, which had multiple confidentiality and integrity issues [149]. However, even after all the enhancements, the IEEE 802.1 1i protocol remained vulnerable to different attacks. In [148], the authors proposed a Smart Grid Secure Protocol (SGSP) solution that creates a more secure node-server connection to achieve DoS resistance. In [150], the authors described a new security protocol that adds authentication and preserves integrity and confidentiality. This protocol controls smart grid transmission lines in a sensor network, and it preserves the network connectivity during node failures.

In addition, the authors of [151] proposed a protocol called scalable, secure transport protocol (SSTP) for smart grid data collection. They stated that integration of transmission control protocol (TCP) with transport layer security (TLS) protocol could provide some scalability issues. Their simulation results based on SSTP showed a high level of security and scalability for smart grid. They also showed that this protocol can reduce memory overhead exponentially. In [152], the authors proposed a lightweight protocol for reliable communication in smart grid network. This protocol solved the security issues of some other protocols, such as IEC 62351 and IEC 61850. Their results show a reduction of the communication cost, solving overhead issues, and improving the privacy and security of data exchange.

In [153], the authors also discussed a countermeasure that uses IPSec and secure neighbor discovery (SEND) protocols. This study focused on using these protocols to prevent any vulnerabilities that may occur in communication with other protocols, such as IPV4 and IPV6. In [154], the authors described compressed transport protocols, such as the datagram transport layer (DTL) in the network layer, which can be considered a good protection mechanism against cyber-attacks in networks. This protocol can mainly protect networks against cyber-attacks that influence data integrity and confidentiality.

Several other studies also introduced and discussed the standards in smart grid as efficient techniques to mitigate the detected cyber-attacks in a network. For example, the author of [155] highlighted some standards, including The US Nuclear Regulatory Commission (NRC) Regulatory Guide (RG) 5.71 (NRC RG 5.71), IEEE Std 1686-2013 IEEE Standard for Intelligent Electronic Devices Cyber Security Capabilities (IEEE 1686), Security Profile for Advanced Metering Infrastructure (security profile for AMI), and the ISO/IEC 27000 series (or ISO27k for short, such as ISO 27001, and ISO 27002) as critical countermeasures in smart grid infrastructure. In general, NRC RG 5.71 is mainly able to establish a comprehensive analysis for computer systems and smart grid networks, identifying the necessary digital assets, and deploying the required security controls. This security measure can mitigate the detected attacks in scope of nuclear infrastructure of smart grid networks.

Another important standard in smart grid is the IEEE 1686, which can provide a complete security control in a network. This standard can mitigate any detected vulnerability in the network. In addition, some other standards, like security profile for AMI, can provide baseline controls for AMI systems in smart grid infrastructures. Also, , ISO27k series standards are critical security controls for managing information through conducting information security management system (ISMS). These standards are capable of identifying the detected threats during the testing and assessment process in the network.

In addition, the authors of [156] described a few other standards, such as IEC 62443 -3-3 and IEEE C37.240. For instance, they discussed in-depth the importance of security controls, such as IEC 62443 -3-3 and IEEE C37.240, which can perform security controls and reliable countermeasures mostly through SCADA systems in the network. Other standards include NISTIR 7628, IEEE 1686, IEEEE C37.240, IEEE 1402, IEC 62056, and ISO/IEC 19790. An overview of essential standards as security controls are provided in Table V.

Table V. List of Important Security Controls in Smart Grid [155-158].

| Security Control | Scope |
| --- | --- |
| NRC RG 5.71 | Security of Nuclear Infrastructure |
| ISO 27001, and ISO 27002 | Security Information System Management |
| Security profile for AMI | Security of AMI |
| IEC 62443 -3-3 | SCADA |
| IEEE C37.240. | Security of Communication Substations |
| NISTIR 7628 | Security of all Components of Smart Grid Infrastructure |
| IEEE 1686 | Security of Vehicular-based Communication Systems in Smart Grid |
| IEEE 1402 | Security of Electric Power Substation in Smart Grid |

| IEC 62056 | Security of meter data exchange in Smart Grid |
| --- | --- |
| ISO/IEC 19790 | security characteristics of cryptographic modules |
| IEC 62351 | Security of communication protocols |
| IEEE 2030 | Smart Grid interoperability for all components |
| IEC 61400-25 | Wind power plant component in smart grid |
| IEEE 1402 | Security of physical and electrical substations |
| IEC 62056-5-3 | Security of AMI component for data exchange |
| ISO/IEC 14543 | Security of home electronic system component |

#### 4.4.1.2. Cryptographic and Authentication

As previously mentioned, one other basic type of countermeasure techniques in smart grid is cryptography and authentication. Most of these techniques aim to protect data integrity, privacy, and confidentiality. In this section, we discuss most common cryptographic functions and authentication methods.

In general, there are two types of cryptographic functions: symmetric and asymmetric functions. In symmetric functions, the encryption and decryption keys are the same or a transformation of one another. The well-known algorithms used as symmetric methods are advanced encryption standard (AES), one-time pad (OTP), and data encryption standard (DES). In asymmetric methods, different keys are used for encryption or decryption: a public key and a private key. The most known asymmetric algorithm is RSA (Rivest, Shamir, and Adleman) [121-122]. In fact, these techniques' efficiency depends on various factors, such as computational resources, processing time, and time complexity. Another technique is elliptic curve cryptosystems. The authors of [92] introduced elliptic curve cryptosystems as a public-key cryptography with the same power as RSA, despite its key size smaller than RSA [123].

Key management techniques play an important role in encryption and authentication approaches. Public key infrastructure (PKI) is a type of key management that guarantees authentication through a network. In [125], the authors discussed some smart grid requirements regarding key management approaches, such as secure management, scalability, evaluability, and efficiency. These requirements must be followed in order to establish a secure key management scheme. Several examples of key management frameworks are provided in [7] as key establishment scheme for SCADA systems (SKE), key management architecture for SCADA systems (SKMA), advanced key management architecture for SCADA (ASKMA), advanced ASKM (ASKMA+), and scalable method of cryptographic key management (SMOCK).

Authentication methods are widely applied in smart grids as countermeasures. In [124], the authors introduced a privacy-preserving data aggregation scheme using authentication methods in the smart grid. Another countermeasure in smart grid is blockchain which is a new emerging technology [126] that can bring considerable advantages to the smart grid's cyber-security. In blockchain technology, distributed structures or ledgers can store digital data without any central authority in peer-to-peer networks. It provides potential solutions as a countermeasure approach, particularly for preventing or eliminating different cyber-attacks, such as man-in-the-middle attacks [127-130] and eavesdropping attacks [128, 130].

#### 4.4.1.3. Intrusion Prevention

Any malicious activity in the network, called intrusion, has to be prevented or eliminated to enhance smart grid performance. One of the traditional methods of preventing attacks on any system is to use firewalls and antivirus. The authors of [131] defined firewalls as a software or hardware systems that can monitor network activities by using several protocols or policies; however, using firewalls and antivirus cannot effectively deal with unknown or sophisticated cyber-attacks. For this purpose, other security techniques, such as network data loss prevention (DLP), intrusion prevention systems (IPS), security information and event management systems (SIEM), File integrity monitoring (FIM), and automated security compliance have been proposed to diminish or prevent the impacts of cyber-attacks on the network.

In general, DLP is a system that can prevent the theft or loss of data through the network, while IPS is an intrusion system that can prevent the identified attacks in the network. IPS and DLP observe the network continuously, identifying malicious activities and abnormalities, and reporting them to the network administrator to prevent them. In [160], the authors evaluated thirty-seven different IPS for smart grid in terms of their architecture, intrusion methodology, and programming characteristics. They also specified that none of these IPS has a self-healing mechanism that they can help during emergencies. In [161], the authors proposed an IPS that mainly protects ZigBee-based home area networks in the smart grid against multiple attack types. Their simulation results demonstrate that this proposed system secures the network against multiple attacks, such as spoofing, eavesdropping, and DoS attacks.

In addition, a few studies used DLP systems as techniques to prevent cyber-attacks in the network. For example, the authors of [162] mentioned DLP as a monitoring system that can diminish the impact of any breach or vulnerabilities in the network; however, this technique usually cannot ensure security for heterogeneous networks such as a smart grid. To address this issue, several other security mechanisms, such as security information and event management systems (SIEM) and automated security compliance, have been proposed to prevent possible intrusions in a network and reduce the risk of cyber-attacks in smart grid. For instance, SIEM can connect security information management (SIM) and security event management (SEM) system. This system constantly analyzes events and provides security alerts if anything unusual occurs [163]. In smart grid, the authors of [164] stated that this system can be used as a good technology to prevent cyber-attacks.

Another practical solution to prevent intrusion in a network is file integrity monitoring (FIM) that prevents any changes in sensitive data and files and determines the possible breaches in the network. In [165], the authors applied the FIM system to protect the integrity of consumers' sensitive data and privacy in smart grid networks. In [166], the authors introduced another technique, called automated security compliance, which is considered an automated tool in the network. The proposed automated security compliance can check through smart grid components in order to guarantee the system configurations are updated. This tool can show a fault in any smart grid component, leading to a security breach in other components. Another practical solution to mitigate the detected intrusion in smart grid network is to sanitize the dataset. For instance, SQL injection attacks mainly happen when a malicious SQL statement is submitted to a web form. To prevent such attacks, sanitizing the dataset can be effective approach in the network [167].

Address space layout randomization (ASLR) is yet another countermeasure in smart grid networks. ALSR is defined as a memory-protection technique for any network against buffer-overflow attacks. These techniques can insert an address space target in any unpredictable locations of the network. Hence, ASLR can reduce or even prevent the risk of memory corruptions in smart grid network. Other simple mitigation techniques for detecting cyber-attacks are web browser extensions (for users), Spam Ware, and moving-target defense. Moving-target defense techniques (MTD) consist of several technologies which are required to increase their security resilience through improving their software diversity. MTD techniques in smart grid are considered as crucial techniques to defer any blended cyber-attacks. For instance, the authors of [168] proposed a MTD-based technique that is considered as a defense technique in order to mitigate detected false data injection attacks in the network. Their simulation results demonstrated that the proposed approach can reduce the attacker's ability to estimate the underlying space model, and that can prevent such cyber-attacks in network.

#### 4.4.1.4. Education

In smart grid, utility services play a significant role in preventing cyber-attacks. For this purpose, the authors of [133] recommended that suitable security training and education for both employees and customers can efficiently prevent the impact of some attacks on networks. For example, tailgating attacks, a common type of social engineering attacks, can be prevented by training professionals. More precisely, employees are supposed to be trained never to give access to any users who do not have badges to access the inside of the utility company, and employees need to have IDs to access to the network [159]. Moreover, employees may be required to discard sensitive data and materials and important files to avoid such attacks.

Some companies implemented security defense frameworks to analyze and mitigate cyber-attacks in their networks. Using these frameworks, they can analyze consumers' profiles to show the existing threats and attacks in smart grid. However, this is not sufficient to minimize the impacts of cyber-attacks. They also need to increase the awareness of employees about cyber-attacks, such as social engineering, and how to prevent them. Another security education is to report security incidents to the IT support team. Reporting incident procedures may help the utility services identify possible vulnerabilities and malicious actions for further reference and avoid that they happen in the future. In addition to employee education, users are responsible for preventing cyber-attacks on smart grid networks. Users must avoid letting someone use their personal ID or password. They also need to be check if they are using legitimate websites before entering any personnel information. Another important venue for cyber-attacks are emails; users as well as employees have to verify that the email is coming from the utility company before clicking on any link embedded in the email [81].

#### 4.4.1.5. Access Control and Cyber-Security Policies

There are a variety of strategies that are effective in managing smart grid networks and determining privileges' access to users and employees. These strategies mainly manage permissions along with providing assurance for an enterprise in a scalable solution. For example, policy-based access control techniques, also known as attribute-based access control, are practical solutions to tackle data security and management. In smart grid infrastructure, authorized employees are required to define some authorization policies in order to give permissions to other employees and users. These policies mostly show the regulations for individuals to provide protection against physical vulnerabilities or cyber-attacks.

Furthermore, integrity checking policy is another method to check if the data has been altered. Therefore, any changes to the network can be observed, and a set of core controls can be applied. Integrity checking in smart grid networks can be performed as a security countermeasure and an indicator of malicious activity. Integrity checking methods can be monitored by authorized employees. Another countermeasure in this category is associated with physical protections. A physical protection is a set of hardware, software, data, and network that protect a network from external or internal vulnerabilities [181] and actions that may cause serious loss or damage to the infrastructure. In addition, a physical protection is an efficient approach in dealing with meter measurement modification attacks that can happen accidentally or intentionally when AMI reports incorrect measurements [167].

Security policies can also improve the security of a smart grid network and reduce the impact of cyber-attacks. These policies are primarily defined by authorized mangers or higher authorities, and they change over time. Such security policies include acceptable user policies, risk assessment standards, personnel security policies, end user key protection controls, and monitoring and logging policies. Although these policies usually focus on providing confidentiality and integrity, they cannot individually guarantee efficient protection in smart grid. Therefore, using several other security controls are considered as necessary steps to secure the network [182].

### 4.4.2. Protective- Based Countermeasures

In this section, we discuss two main categories of protective-based countermeasures, namely spread spectrum and game theory-based techniques.

#### 4.4.2.1. Spread Spectrum Techniques

In smart grid, spread spectrum techniques are defined as a major approach in which a generated signal with specific bandwidth is deliberately spread in a frequency domain leading to a wider bandwidth. Spread spectrum techniques are known as effective techniques to prevent jamming attacks in networks. These techniques can be divided into frequency hopping (FHSS) and direct sequence (DSSS). In the following, we briefly describe these two types in the scope of smart grid infrastructure.

In FHSS techniques, signals are transmitted by changing a carrier frequency among several distinct occupied frequencies. In [169, 170], the authors introduced an FHSS technique to provide protection against jamming and collision attacks in the network. Their results showed that the total required bandwidth of this technique is wider than similar data with a single carrier frequency. In [60], the authors mentioned the advantages of FHSS techniques as countermeasures, including dealing effectively with the multipath effect.

DSSS techniques are used to decrease the overall signal interference. The direct sequence creates the transmitted signals much higher than the information signals. For example, in [171], the authors used a DHSS-based approach to mitigate the detected jamming attacks in the network. In this study, the authors conducted their approach according to the dynamic tree-based scheme; however, it generates a huge maintenance overhead. Although the proposed method provided good protection against jamming attacks at the physical layer, this technique required very expensive computational resources. In [172], the authors also investigated the use of a DSSS technique by applying code division multiple access (CDMA) to prevent jamming attacks in the network. In [173], the authors recommended a hybrid approach that combines FHSS and DSSS to protect the network against jamming attacks. The authors compared their results with FHSS and DSSS, and concluded that a hybrid of FHSS/DSSS can provide a low probability of detection, low probability of interception, and improvement of the ability to deal with near far problems.

#### 4.4.2.2. Game Theory-Based Techniques

Game theory-based techniques are considered as mathematical models that have strategic interactions among rational decision makers [174]. In [175], the authors proposed a two-layer game theory prevention technique for false data injection attacks in smart grid. In this study, their proposed approach used data from multiple sources in order to increase the prevention rates of attacks. So, they developed a zero-sum static game theory that optimize the deployment of various defense resources. The authors of [176] also proposed a game theory model based on the minimax regret method. This multi-level game theoretic framework provides a cost-effective and computationally efficient approach for large-scale power systems and smart grid infrastructure.

In [177], the authors introduced an approach based on game theory for defending against cyber-attacks in smart grid. In this work, they applied a game theory-based method which can identify cyber-attacks for smart energy scheduling of smart grid. The authors of [178] also designed a game theory approach to prevent against dynamic cyber-attacks in smart grid networks. Their model strategically identifies the chronological order of cyber-attacks that can occur, then protects the network against these attacks. Their simulation results proved that the proposed model is good and effective in detecting cyber-attacks.

The authors of [179] developed a game theory-based technique for smart grid, which according to their results, is 4.5 times faster than other existing studies. This technique also achieves a low communication and storage cost. In [180], the authors also proposed a system, using dynamic game theory technique, as countermeasure that analyzes the attacks in cyber-physical system of the network. They mostly used a hybrid model that combines particle swarm optimization technique, game theory, and sequential quadratic programming technique to validate their model.

*4.5. Comparison and Discussion*

Each detection and countermeasure category has some advantages and disadvantages. In this section, we briefly describe these benefits and shortcomings of these techniques as shown in Table V and VI. As we discussed earlier, our proposed classification has different categories, namely localization-based, AI-based, prediction models, and intrusion detection systems. A comprehensive summary of advantages and disadvantages of these techniques are provided in Table V. As shown in Table V, localization-based techniques, the first category of the detection classification, have some limitations, such as high processing time and synchronization. In addition, these techniques present lower complexity, and having the locations of the malicious users is not always necessary for detecting attacks.

The second category, AI-based techniques, can provide a high level of detecting cyber-attacks in smart grid networks; however, these techniques, such as machine learning, deep learning, data mining, and fuzzy logic methods, have a very high implementation cost for smart grid networks. Nevertheless, their detection rates are high. Furthermore, these techniques mostly provide a low false alarm rate in risky situations and are also considered as the fastest techniques in detecting cyber-attacks. It is worth to mention that AI-based methods require a proper dataset to test and implement their algorithms; however, due to security reasons, working with real data may not be possible in smart grid networks.

The third category of our detection classification, prediction models, can generally provide a better knowledge about the trend of an attack occurrence, while these techniques have to be used along with proper and high-quality data. In fact, incorrect and low-quality data may lead to a poor performance of these models. In addition, using prediction models is one of the main keys of identifying future security risks and attack incidents. Therefore, they can be an effective model in detecting cyber-attacks in smart grids.

One of the most robust and flexible detection methods in smart grid is filtering-based techniques, which usually have high computational complexity. These techniques are also simple to develop and their detection rates are high. It is also worth to mention that these techniques are very cost effective. Despite all of their benefits, they require a fixed threshold, which can be challenging to select.

The last category of our detection classification presents intrusion detection systems, which are widely used to detect multi-step cyber-attacks in the network. These systems are usually used along with other techniques, such as AI-based methods. In general, in smart grid networks, there is no need for an authorized user to control the intrusion detection systems, and they mostly perform without any centralized authorizations. Also, these techniques have high rates of false alarm and require high memory storage. Therefore, these systems still need more development and improvement to detect complicated attacks in the smart grid.

Table V. A Discussion of Advantages and Disadvantages Of Different Detection Techniques in Smart Grids.

| Detection Technique | Advantage | Disadvantage |
|---|---|---|
| Localization-based | Less complexity. Always malicious users' location is not needed. | Needs synchronization. High processing time. |
| AI-based | Usually high detection rate and low false alarm. | For learning process, a proper dataset for training and testing is required. |
| Prediction models | Analysis of the current and historical data. Understanding a better trend. Identify potential future risks and opportunities. | Incomplete and poor data quality lead to inaccurate results. |

| Filtering-based | Easy implementation. Robust. | Usually fixed threshold. Usually high computational complexity. |
| Intrusion Detection System | No need to be centralized. | High false rate. High memory storage. |

As presented in Table VI, we compare the proposed countermeasure categories to discuss their advantages and disadvantages. According to this table, secure protocols and standards provide flexible solutions to prevent cyber-attacks in the smart grid. In addition, these methods are simple to manage and maintain; however, there are still no protocols in the smart grid infrastructure that guarantee a high level of security. Moreover, secure protocols and standards deal with a limited frequency of communication, which may lead to the lack of performance in the network.

Cryptographic and authentication techniques are also used in confidential scenarios, although their implementation complexity is high and they are an inefficient solution. Another preventive countermeasure, as shown in Table VI, is intrusion prevention, which can protect the privacy of the network and users while preventing abnormal activities. Furthermore, to provide a higher security and to mitigate the detected attacks, intrusion prevention methods are not highly recommended to be used without any other techniques. Education techniques is also another category in our countermeasure classification, which is considered as a simple and easy to understand method for users. Only educating users, however, is not enough to guarantee any attack prevention. Access control and cyber-security policies also are considered as another prevention approach that have a high scalability, which is easy to understand. Furthermore, this approach cannot guarantee the protection against cyber-attacks and is only compatible with small-scale networks.

Game theory-based techniques also protect the smart grid against multiple cyber-attacks with optimal solutions and high data rates; however, in these methods, it is necessary to have mobile users to develop such techniques. It is clear that, in some scenarios, it is impossible to have mobile users. Therefore, game-based countermeasures cannot be used in all conditions. Another protective countermeasure is spread spectrum techniques, which provide high levels of protection. However, they may have some limitations, such as a complicated implementation for smart grids and an inefficient bandwidth. Despite these challenges, spread spectrum techniques can be a good solution to mitigate the detected cyber-attacks in smart grids.

Table VI. A Summary of Advantages and Disadvantages of Countermeasure Techniques in Smart Grids.

| Countermeasure methods | Advantages | Disadvantages |
|---|---|---|
| Secure protocols and standards | Flexibility. Simple to manage and maintenance. | No high secure protocols in smart grid. Limited frequency of communication. |
| Cryptographic and authentication | Cryptographic algorithms benefits. Confidentiality is recommended. | Implementation complexity. Not always efficient. |
| Intrusion prevention | Privacy protection. Prevent abnormal network activities. | Needs to implement with other countermeasure techniques to be able to prevent attacks. |
| Spread spectrum techniques | High level of protection. | Complicated implementations. Inefficient bandwidth. |
| Game theory-based | Optimal solution. High rates of data. | Mobile users are necessary. |
| Education | Simple. Provide enough knowledge to users and employees. | Not enough to protect and prevent the network against attacks. |
| Access control and cyber-security policies | High scalability. Simple to understand. | Not ensure protection against attacks. Suitable for small-scale networks. |

V. Challenges and Future Directions

With smart grid deployment, this technology is exposed to several cyber-attacks like any other heterogeneous system. It is found that there is a plethora of challenges in the security of the smart grid in order to provide a reliable, secure, and protective framework. Moreover, there are still several challenges and open questions about the security of the smart grid that need to be addressed and answered. Also, most smart grid advancements are in the early stage and they are considered more conceptual rather than practical. Therefore, studying challenges and future directions play an important role in the advancement of the smart grid.

For example, the increased number of connected devices with unsecure protocols makes smart grid networks vulnerable to new attacks. Each device connected to the network can be considered as a possible point of entry. There are numerous studies that aimed to enhance the security of the smart grid; however, some of the existing techniques have fundamental limitations, and there are still a number of challenges to address. For instance, existing

IDSs still deal with some limitations, such as a low detection accuracy and a high false positive rate. Several studies used machine-learning techniques with these systems to improve their performance. However, machine learning models require large datasets which are not widely shared by researchers. A few groups share their datasets, and these datasets do not include data gathered from real attacks. Therefore, there is a need for developing and sharing datasets for machine learning training and validation.

Cryptographic and authentication techniques barely support AMI and WAN entities. Several techniques have been proposed on this topic; however, these techniques are only compatible with SCADA systems. Thus, one research direction is to develop key management techniques specific to AMI and WAN components. For detection techniques that use artificial intelligence, the models have to be extensively trained before any cyber-attack happens. Therefore, there is a need for techniques that classify not only incoming signals, but also prevent new attacks and help the system recover from them. Another issue is the fact that existing techniques deal only with one attack. These techniques are inefficient in detecting complex and distributed attacks. Thus, there is a need for layered frameworks that can prevent, detect, and mitigate cyber-attacks in smart grid infrastructure. Regarding current protocols for smart grids, their main purpose is connectivity, not security. None of these protocols provide a high level of security. The confidentiality, privacy, integrity, and accountability can easily be violated with such existing protocols. Therefore, there is a need for new secure protocols for smart grid networks.

## VI. Conclusion

The security of smart grid networks is of paramount importance and plays a pivotal role in the implementation of smart grid systems. However, prior studies have shown a constrained role in evaluating cyber-security solutions for smart grid networks. Therefore, this paper considers the shortcomings of the existing surveys and provides an in-depth description of potential attacks that target smart grids and an evaluation of different security solutions. In this paper, we propose a benchmarking of cyber-attacks in terms of the integrity, availability, confidentiality, and accountability and a classification based on OSI communication layers. Moreover, we present a new classification for the existing detection techniques, which is mainly divided into localization-based, AI-based, prediction models, filtering techniques, and intrusion detection systems. We also classify the countermeasure techniques into preventive and protective techniques. In the preventive countermeasures, we describe secure protocols and standards, cryptographic and authentication, intrusion prevention, education, access control, and required cyber-security policies approaches. For the protective countermeasure category, we discuss spread spectrum techniques and game theory in the smart grid. Finally, we describe the existing challenges that can guide future research directions. This survey has highlighted the requirements of new solutions, which can collectively resolve the problems related to security challenges in the smart grid infrastructures without compromising the performance and functionalities of this type of network.